\documentclass{article}
\usepackage[preprint]{colm2026_conference}

\usepackage{booktabs}
\usepackage{float}
\usepackage{graphicx}
\usepackage{hyperref}
\usepackage{microtype}
\usepackage{url}
\usepackage{amsmath}
\usepackage{amssymb}
\usepackage{tikz}
\usepackage{lineno}

\definecolor{darkblue}{rgb}{0,0,0.5}
\hypersetup{
  colorlinks=true,
  citecolor=darkblue,
  linkcolor=darkblue,
  urlcolor=darkblue,
  pdftitle={Authorization Before Context: A Model-Neutral Audience Boundary Against Cross-Audience Memory Leakage in Agentic Systems},
  pdfauthor={Sibo Liu},
  pdfsubject={Security and authorization for shared memory in agentic systems},
  pdfkeywords={AI agents, LLM security, agent memory, access control, contextual integrity}
}

\title{Authorization Before Context:\\ A Model-Neutral Audience Boundary Against\\ Cross-Audience Memory Leakage in Agentic Systems}

\author{
Sibo Liu\\
Independent Researcher\\
\href{mailto:abc-paper@outlook.com}{\texttt{abc-paper@outlook.com}}
}

\begin{document}

\raggedbottom

\ifcolmsubmission
\linenumbers
\fi

\maketitle
\lhead{Author preprint.}

\begin{center}
\small
Accepted for presentation at AdvML-Frontiers x CoTMA,
a non-archival workshop at COLM 2026.
\end{center}

\begin{abstract}
A personal language agent learns a fact from one audience and may later place it in the prompt it assembles for another. This memory-to-context step is an attack surface: ambiguous or inconsistent channels, cross-audience prying, and poisoned memory can each cause the system to assemble context containing a fact relevant to the query yet unauthorized for the current viewers. We introduce \emph{authorization before context}: a single, anti-monotone audience-membership rule applied at the memory-to-context transition. Each item carries the audience present when it was recorded; the current viewer set is read from channel metadata and falls back to public when ambiguous; and the item is admitted only when every current viewer already belonged to its audience. We prove that this rule gives every participant cross-channel recall while ensuring, \emph{by exclusion rather than by model behavior}, that nothing recorded for a narrower audience reaches a broader one and that poisoned memory cannot widen its own audience. The boundary is a \emph{model-neutral} invariant on the exact assembled context: a forbidden fact must be absent before the model is called. On a synthetic Contextual-Integrity suite, no forbidden fact entered the context our boundary assembled, whereas unscoped baselines included such facts by construction; we further audit that every read path fails closed. The evidence is preliminary and synthetic.
\end{abstract}

\section{Introduction}

Persistent memory shifts the security boundary of a personal language agent. Such an agent is a delegated proxy for its owner, much like a human executive assistant. It can hold private conversations with the owner and, on the owner's behalf, participate in one-to-one and group conversations with external contacts. Each prompt it assembles draws on sources such as recent messages, long-term memory, and retrieved knowledge~\citep{packer2024memgpt,chhikara2025mem0,rasmussen2025zep}. The resulting threat is compositional and arises at the interaction layer: a fact learned in one conversation can be summarized, stored, retrieved on a later turn, and then included in the prompt assembled for a \emph{different} audience. When a member of that later audience is itself an agent, this cross-audience exposure is a memory-layer instance of inter-agent trust exploitation. Memory-augmented agents are known to leak private information or to be compromised through shared memory, retrieved experience, poisoned records, and tool arguments~\citep{el_yagoubi2026agentleak,chen2024agentpoison,srivastava2025memorygraft,pulipaka2026hiddenmemory,zhang2026memmorph,wang2026mempoison}. The principle we adopt is that memory should follow its \emph{audience}, not the channel that carries it.

Three common designs are insufficient. (i)~\emph{Relevance-only retrieval} ranks memory by how well it matches the query and is blind to who may see an item on the current turn~\citep{packer2024memgpt,chhikara2025mem0}. (ii)~\emph{Output filtering} acts only after the prompt has been built and read by the model, so the unauthorized fact is already in its context; redacting the response cannot undo that exposure, and scoring the answer cannot detect it, because the model can answer acceptably even on forbidden input~\citep{mireshghallah2025cimemories}. (iii)~\emph{Product-level isolation} side-steps the question instead of answering it. Some frameworks silo memory by user, agent, session, or namespace~\citep{langchain2026langgraphmemory,mem02026entityscopedmemory,zep2026concepts,letta2026statefulagents}, and self-hosted personal agents keep one global store but use an allowlist to gate who may converse~\citep{openclaw2026,hermesagent2026}. Principal-scoped retrieval is a finer variant that tags memory by user and filters reads to that user~\citep{langchain2026langgraphmemory,mem02026entityscopedmemory}. But a user can belong to several audiences. A per-user tag cannot distinguish one-to-one facts from group facts involving the same user. None of these systems uses the audience (those present at recording time) as the unit, so the core question stays open: when a durable fact is visible to exactly some participants, which later prompts may include it?

We answer with an authorization boundary at the memory-to-context transition that addresses the three gaps directly. Against relevance-only retrieval, it admits by \emph{audience membership} rather than by relevance. Against output filtering, it decides admission \emph{before} the prompt is assembled, so no unauthorized fact ever reaches the model. Against per-silo memory, it keeps one shared store yet confines each item to its audience, so participants still recall their own facts across channels. Mechanically, an item's audience is fixed at record time from the authenticated membership of its source conversation and re-checked at assembly: an item is admitted only when every current viewer already belonged to its audience, with the viewer set read from channel metadata and resolved to public when ambiguous. Admission is decided by membership rather than by a viewer's type, so human and agent viewers are treated alike. Because audiences are sets, the check is a Zanzibar-style relationship test~\citep{pang2019zanzibar,cutler2024cedar} grounded in Contextual Integrity~\citep{nissenbaum2004contextualintegrity}, moved from object access to prompt-context assembly.

To our knowledge, no prior memory system combines these three properties: an admission decision based on the \emph{audience membership} of a turn's participants, with the viewer set read from the transport rather than supplied as an authoritative policy, and resolved fail-closed; a single rule that is at once recall-preserving and one-way confining; and a model-neutral check on the exact assembled context. Filtering reads before assembly is not itself new~\citep{rezazadeh2025collaborativememory}; the contribution is this basis and these guarantees in combination.

Concretely, we contribute: (1)~an audience-membership boundary whose single admission rule we \emph{prove} yields both per-participant cross-channel recall and one-way confidentiality; (2)~a portable, model-neutral invariant on the exact assembled context, enforced by exclusion rather than by redaction or output scoring; (3)~implementation-grounded authorization traces on a synthetic suite, in which a single audience decision governs three structurally different memory stores, together with a read-path fail-closed audit and an internal, human-reviewed live-verification process; and (4)~a disclosure-safe artifact and a positioning against access-controlled-memory systems and contextual-integrity benchmarks. We claim only that, in the evaluated configuration, no known forbidden fact entered unauthorized context and every audited read path failed closed; action safety, audience widening, and memory quality are out of scope or future work.

\section{Threat model}

We consider a personal agent that serves one owner across private and shared channels. The assets to protect are the information exchanged in each conversation, whether owner-private, one-to-one, or group; any memory the agent derives from it; the exact context assembled for each model call; and the audience labels and provenance that justify what is included. We assume three threats: \emph{cross-audience prying}, in which participants elicit facts from a scope they do not belong to; \emph{channel ambiguity}, in which an inconsistent participant set leads a naive system to over-share; and \emph{poisoned memory}, in which content planted in one scope later reaches a viewer outside that scope~\citep{chen2024agentpoison,srivastava2025memorygraft,pulipaka2026hiddenmemory}. These map to sensitive-information disclosure, broken access control, and memory/RAG poisoning in current taxonomies~\citep{owasp2026agentic,he2026attriguard,cui2026spore,huang2026observablechannels}.

\begin{figure}[t]
\centering
\vspace{-2.0em}
\includegraphics[width=0.72\linewidth]{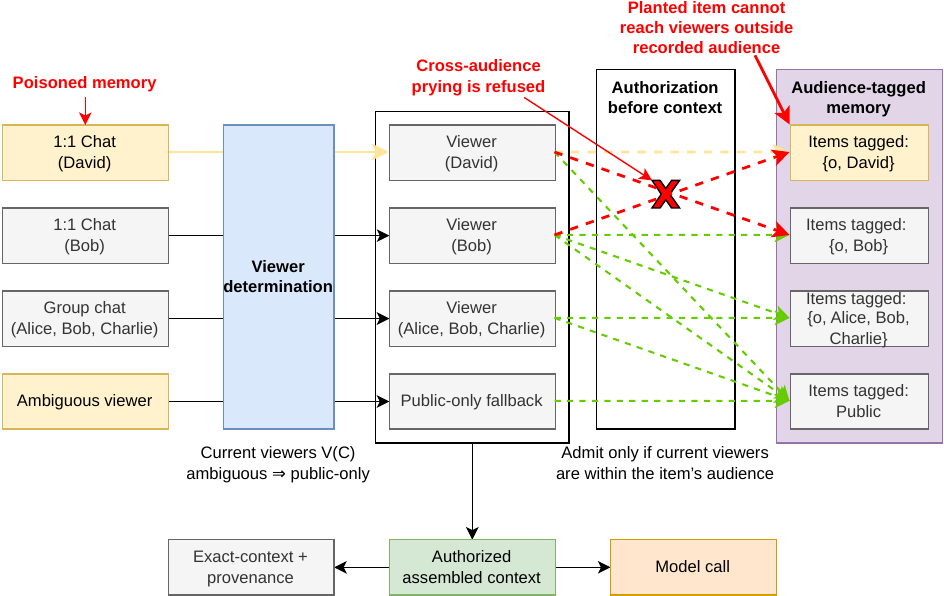}
\caption{Authorization before context. Channel metadata resolves current viewers, with ambiguity falling back to public-only context. Audience-tagged items are admitted only when all current viewers are within the recorded audience; prying is refused, poisoned memory cannot reach outside that audience, and exact-context provenance records the admitted context.}
\vspace{-1.2em}
\label{fig:threat-dataflow}
\end{figure}

The trust boundary we study is the point where stored memory is assembled into a turn's prompt context (Figure~\ref{fig:threat-dataflow}). Once an unauthorized item is in that context, later defenses are too late: the model has already read it. We therefore make the boundary checkable \emph{before} the model is called. Its guarantees are deliberately narrow: retrieval fails closed when the viewer set is uncertain; admission is by audience membership, not content redaction; and the invariant is model-neutral, failing whenever a forbidden fact enters the context regardless of what the model then says. This paper addresses that read boundary only. We assume the transport correctly names a conversation's participants, and we do not attempt a general defense against prompt injection; the audience rule instead \emph{contains} planted content, limiting any poisoned item to viewers already within the audience it was recorded for (property P4 below).

\section{The audience-membership boundary}

In our design, all memory lives in one shared, \emph{audience-tagged} store: several structurally different stores hold the items, but a single audience decision governs them as one. Each \emph{memory item}, a fact the agent has recorded from a conversation, carries the audience that was present when it was recorded. An audience is identified by its resolved member set, not by the channel or thread it is observed on; two turns share a scope whenever their members coincide, even across different channels or threads. The \emph{public} audience is a reserved value that denotes all participants and is the default for information any viewer may see. A single conversation turn defines a \emph{container} $C$ with a set of current viewers $V(C)$; before the agent assembles the prompt it will send on that turn, each candidate memory item is authorized against $V(C)$. A memory item $F$ with audience $\mathrm{Aud}(F)$ is admissible iff
\[
V(C)\subseteq \mathrm{Aud}(F),
\]
that is, every current viewer already belonged to $F$'s audience. The viewer set is read from channel metadata (sender, container, observed participants); when that evidence is missing or inconsistent, it is treated as unknown and authorization fails closed to public-only. Admission is by exclusion: an unauthorized item is never placed in the context, independent of any model.

\paragraph{Formal properties.}
Let $U$ be the set of participants, with a distinguished owner $o\in U$, and let $M$ be the set of memory items. Each item $m\in M$ has an audience $\mathrm{Aud}(m)\subseteq U$ that contains the owner ($o\in\mathrm{Aud}(m)$); the \emph{public} items are those whose audience is everyone, $\mathrm{Pub}=\{\,m\in M : \mathrm{Aud}(m)=U\,\}$. A container $C$ (one conversation turn) has a current-viewer set $V(C)\subseteq U$ and admits
\[
  \mathrm{Adm}(C)=\{\,m\in M : V(C)\subseteq\mathrm{Aud}(m)\,\}.
\]
Public items are admitted automatically, since $V(C)\subseteq U=\mathrm{Aud}(m)$. When $V(C)$ cannot be determined, the fail-closed rule sets $V(C)=U$, so $\mathrm{Adm}(C)=\mathrm{Pub}$. Four properties follow.

\textit{P1 (one-way confinement).} By the rule, admission entitles every current viewer:
\[
  m\in\mathrm{Adm}(C)\ \Longrightarrow\ V(C)\subseteq\mathrm{Aud}(m).
\]
Hence an item never reaches a viewer outside its audience, and nothing recorded for a narrower audience reaches a broader one.

\begin{figure}[t]
\centering
\begin{tikzpicture}[font=\small]
  \node (g) at (0,1.7) {$\{o,B,A\}$ (group)};
  \node (b) at (-2.7,0.6) {$\{o,B\}$ (Bob 1:1)};
  \node (a) at (2.7,0.6) {$\{o,A\}$ (Alice 1:1)};
  \node (o) at (0,-0.5) {$\{o\}$ (owner)};
  \draw[->] (g) -- (b);
  \draw[->] (g) -- (a);
  \draw[->] (b) -- (o);
  \draw[->] (a) -- (o);
\end{tikzpicture}
\caption{Audience lattice. A fact with audience $A$ enters a container $C$ iff $V(C)\subseteq A$: admission flows \emph{downward} (arrows) to smaller current audiences and never upward. Bob's one-to-one ($\{o,B\}$) admits a group fact ($\{o,B,A\}$), but the group never admits a fact whose audience is $\{o,B\}$; the owner ($\{o\}$) is at the bottom and sees the union.}
\label{fig:lattice}
\end{figure}
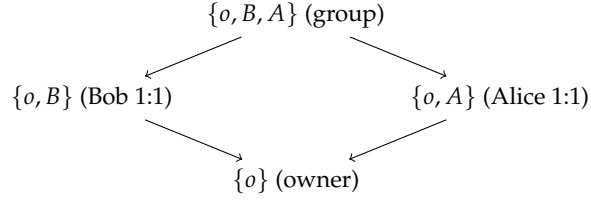

\textit{P2 (anti-monotonicity).} A smaller current audience admits at least as much,
\[
  V(C_1)\subseteq V(C_2)\ \Longrightarrow\ \mathrm{Adm}(C_2)\subseteq\mathrm{Adm}(C_1).
\]
A one-to-one with a contact $c$, where $V(C)=\{o,c\}$, thus recalls every item whose audience includes $c$, across all channels; a group never surfaces a fact whose audience excludes a member; and the owner, where $V(C)=\{o\}$, sees the union of all items (Figure~\ref{fig:lattice}). A single rule yields both: confinement (P1) within a container, recall (P2) across containers.

\textit{P3 (fail-closed soundness).} Under ambiguity $V(C)=U$, so $\mathrm{Adm}(C)=\mathrm{Pub}$ and no non-public item is admitted.

\textit{P4 (poisoning containment).} Write-time binding fixes an injected item's audience to its origin container's authenticated membership $A$, independent of the item's content: $\mathrm{Aud}(m')=A$. By P1,
\[
  m'\in\mathrm{Adm}(C)\ \Longrightarrow\ V(C)\subseteq A,
\]
so a planted item reaches only viewers already within $A$ and cannot enlarge its own audience. P4 is a confidentiality property, not an integrity guarantee: planted content may still reach viewers already within its recorded audience, including the owner, but it cannot cause an unauthorized memory item to enter the assembled context.

These properties hold for the \emph{rule}; Section~\ref{sec:results} checks that the \emph{implementation} realizes them.

We keep the raw record, derived memory, and exact prompt as distinct layers linked by provenance, so that each admission is measurable: an exact-context snapshot shows what entered the prompt, a provenance link ties each item to its source and audience, and an authorization trace records why it was admitted or why the turn fell back to public-only (Figure~\ref{fig:threat-dataflow}). Because the snapshot is captured at the model-request boundary, the inclusion invariant is checked against exactly what is assembled rather than a reconstruction, and the same capture point makes the boundary auditable on live traffic, not only in the benchmark.

% Legacy architecture-planes figure removed: redundant with the data-flow and
% lattice figures, over-detailed relative to the claim, and de-anonymization risk.

\section{Evaluation protocol}

Evaluation uses a fixed-seed synthetic generator that assigns Contextual-Integrity labels (subject, sender, recipient, information type, transmission principle) to facts and audiences~\citep{nissenbaum2004contextualintegrity}; no real messages are used. The evaluated configuration runs with scope authorization active on every read path. The cited report uses seed 42617 and 79 scenarios across 13 families spanning owner-private, exact-shared, directional cross-channel, fail-closed, adversarial, and poisoning settings. The boundary metric is \emph{unauthorized context inclusion}, the number of forbidden facts in the exact-context artifact, counted before any answer. We also report the agreement between the observed authorization decision and the expected audience policy, the fail-closed coverage, and the permitted-context coverage. Provenance, decision, and source-plane labels are not a coverage result but a guarantee of the artifact schema: the architecture attaches them to every admitted item, making each admission auditable.

Baselines are definitional projections over the same scenarios, not tuned runtimes: an \emph{unscoped} projection returns all expected context and therefore includes the forbidden facts by construction, marking the upper bound on leakage that the audience rule must drive to zero. Beyond the synthetic suite, we audit that every prompt-context read path resolves authorization before assembly. The reported results rest on sealed synthetic and redacted artifacts whose digests support independent verification without access to production source. Appendices~\ref{app:repro-artifacts}--\ref{app:worked-example} present the sealed run, the claim-to-evidence closure, and one end-to-end synthetic scenario.

\section{Results}
\label{sec:results}

\begin{table}[t]
\centering
\small
\begin{tabular}{lrrl}
\toprule
Adversarial setting & Scen. & Forbidden incl. & Fail-closed \\
\midrule
owner-private (private$\to$shared pressure) & 8 & 0 & -- \\
exact shared (one-to-one, group) & 16 & 0 & -- \\
directional cross-channel & 4 & 0 & -- \\
trusted canonical shared & 4 & 0 & 4/4 \\
ambiguous / unknown viewer & 6 & 0 & 6/6 \\
mixed-inbound (inconsistent evidence) & 6 & 0 & 6/6 \\
adversarial boundary & 8 & 0 & -- \\
negative control & 10 & 0 & -- \\
other (public, write-admission, temporal, optional) & 17 & 0 & -- \\
\midrule
\textbf{all 79 scenarios} & \textbf{79} & \textbf{0} & \textbf{16/16} \\
\bottomrule
\end{tabular}
\caption{Red-team results by setting (cited report \texttt{run\_1780535616\_e37d6e1c}, rev.\ \texttt{7452d12b}); the rows partition all 79 scenarios from the full 13-family suite, the final row grouping families that have no fail-closed cases. Model-neutral assertions pass 253/253, the per-query conjunction of these decision checks, not a separate measurement.}
\label{tab:primary-results}
\end{table}

No forbidden fact entered unauthorized context under any adversarial setting: 0 of 79 scenarios (Table~\ref{tab:primary-results}). A hermetic run that exercises the real read path without a model reproduces this 0/79, so the boundary metric is observed on the assembly path, not only projected. The observed authorization decision matched the expected audience policy in all 79 scenarios, fail-closed coverage was 16/16, and permitted-context coverage was 79/79. The adversarial-boundary and write-admission families, covering planted-content and storage-binding cases, exercise P4: an item recorded under one audience never surfaced in a context containing a viewer outside its origin audience, consistent with write-time binding of its audience to the origin membership; an indeterminate source membership is refused rather than bound to a guessed audience.

We then audited every retrieval path that can place memory into a prompt. A single audience decision, derived from the current viewers, governs three structurally different stores: an in-turn recency buffer, a summarized long-term store, and a knowledge graph, each enforcing it at a different point. Recent-message recall is confined to the current turn; the summarized long-term reads push the authorized scopes into the query, so only admissible items are fetched and the assembled context is filled with authorized facts rather than thinned by later removal; and graph retrieval, comprising hybrid search and direct entity lookup, alone fetches first and then drops any item whose audience cannot be checked. The direct-lookup primitive enforces the audience check internally, filtering its results against the authorized scopes before returning, so the guarantee holds by construction rather than by caller convention; it returns an identical \emph{not-found} response whether an entity is absent or merely unauthorized, an existence opacity. Every path is therefore fail-closed, realizing P1 and P3. One surface resists per-item authorization: a knowledge-graph entity summary is an aggregate that may span audiences and carries no per-item provenance, so it cannot be checked fact-by-fact; we therefore fail closed here as well, withholding such summaries from non-owner readers rather than admitting content we cannot authorize.

Beyond the synthetic suite, an internal, human-reviewed live-regression fixture captured the exact assembled context across repeated primary and fallback runs of a real, pinned model, and its row-by-row comparison matched the human-approved baseline; we hold this evidence internally, but the verification records (approvals and content digests) are releasable after review.

\section{Related work and limitations}

The closest prior system is Collaborative Memory~\citep{rezazadeh2025collaborativememory}, which also keeps a shared, provenance-tracked store and filters views before assembly. We differ on three axes. \emph{Access unit}: ours is the audience membership of conversation participants rather than a capability graph over agents and resources; this membership basis yields the anti-monotone recall and one-way confinement proved above. \emph{Viewer evidence}: ours is read from the transport and resolved fail-closed, whereas their permission graph is taken as authoritative input. \emph{Enforcement}: ours admits by exclusion and is judged by a model-neutral inclusion invariant, whereas theirs redacts fragments and is judged by task utility.

CIMemories~\citep{mireshghallah2025cimemories} benchmarks contextual integrity of persistent memory but scores violations in model \emph{output}; AgentLeak, MAGPIE, memory extraction, and channel-leakage work likewise audit outputs or recovered data~\citep{el_yagoubi2026agentleak,juneja2025magpie,wang2025unveilingprivacy,cui2026spore,huang2026observablechannels}. We instead check the assembled context before model output, preventing leaks by construction. Memory benchmarks and systems study multi-session tasks, organization, and durable recall~\citep{he2026memoryarena,shutova2026structmemeval,hu2026memoryagentbench,yang2026groupmembench,xu2026memgym,packer2024memgpt,chhikara2025mem0,zhong2023memorybank,rasmussen2025zep}; we make no recall, isolation, or answer-quality superiority claim. Agent-memory security surveys find confidentiality underexplored relative to integrity and poisoning~\citep{lin2026mnemonic}; our work targets that gap. Self-hosted personal agents such as OpenClaw~\citep{openclaw2026} and Hermes Agent~\citep{hermesagent2026} authorize \emph{who may converse} and serve those conversations from one global memory; we instead admit each item only to audiences that already contained it. Zanzibar, Cedar, and Contextual Integrity supply the authorization-before-access and information-flow vocabulary we adapt~\citep{pang2019zanzibar,cutler2024cedar,nissenbaum2004contextualintegrity}.

The data are synthetic. Because the boundary is a deterministic membership test rather than a model behavior, its zero-leakage result holds by construction rather than by measured attack success. Audience \emph{widening}, the promotion of facts that several contacts independently know into a shared audience, is the recall-completeness dual of leakage and may be worth exploring in future work. We deliberately accept overblocking rather than infer such unions, because widening would re-introduce content inference, and thus leakage risk, into an otherwise content-free boundary; doing it safely would require inferring access relationships beyond the static tuples that authorization systems assume. We make no production-reliability, latency, user-study, deletion, conflict-resolution, action-safety, or memory-quality claims; we do not release the production system or its source, though the synthetic benchmark, evaluator, schemas, and redacted artifacts are disclosure-safe and releasable, and the live-verification process is internal operational support rather than a public live-efficacy claim.

\paragraph{Future work.} Three directions remain. First, coverage should deepen by driving the same audience decision through every store we claim, including graph-backed retrieval, which this hermetic run audits as a read path but does not yet exercise as seeded benchmark material. Second, write admission should be evaluated as a measured boundary, generalizing P4's containment of planted content from read-time exclusion to record-time binding. Third, evaluation should broaden to a larger adversarial suite spanning forged metadata, stale labels, multi-hop transitions, and overlapping audiences, with executable baselines that expose which boundary each alternative lacks rather than serve as a leakage leaderboard.

\typeout{===MAINBODY-ENDS-ON-PAGE \thepage===}
\bibliography{references}
\bibliographystyle{colm2026_conference}

\section*{Ethics, reproducibility, and AI-use disclosure}

This paper evaluates synthetic or redacted boundary artifacts and does not publish private messages, prompt bodies, provider settings, or cloneable deployment details. The cited report is generated from fixture-seeded benchmark runs and records source revision, command, artifact digests, disclosure levels, and replay metadata; live-verification artifacts are used as operational and accountability support, not as disclosure of private live traces. All artifacts shown in the appendices are synthetic or redacted excerpts of the sealed evidence package.

AI disclosure: this manuscript was prepared with assistance from AI-powered writing and coding tools for drafting, revision, citation-format cleanup, and figure integration. The author directed the claims, evidence selection, and limitations, and remains responsible for the accuracy of all text, citations, figures, and conclusions, which require author review before submission.

\appendix

\section{Related-work comparison}

Table~\ref{tab:related-work-dimensions} summarizes how the audience-membership boundary relates to the closest prior systems and authorization frameworks, without exposing private implementation detail.

\begin{table}[H]
\centering
\small
\begin{tabular}{p{0.24\linewidth}p{0.32\linewidth}p{0.32\linewidth}}
\toprule
Comparison area & Representative work & Boundary relative to this paper \\
\midrule
Access-controlled shared memory & Collaborative Memory & Capability graph + utility eval; ours is audience membership + exclusion + model-neutral inclusion invariant. \\
Contextual-integrity / leakage & CIMemories, AgentLeak, MAGPIE & Output- or extraction-scored; ours is checked on assembled context before scoring. \\
Memory benchmarks & MemoryArena, StructMemEval, MemoryAgentBench, GroupMemBench, MemGym & Task/organization/recall; ours is an audience-boundary invariant. \\
Memory systems & MemGPT, Mem0, MemoryBank, Zep & Durable memory; we make no backend-superiority claim. \\
Self-hosted personal agents & OpenClaw, Hermes Agent & Allowlist gating of who may converse, then one global memory; ours adds per-item audience confinement. \\
Authorization / privacy models & Zanzibar, Cedar, Contextual Integrity & Vocabulary we adapt to prompt-context assembly. \\
\bottomrule
\end{tabular}
\caption{Related-work comparison dimensions.}
\label{tab:related-work-dimensions}
\end{table}

\clearpage

\section{Reproducibility and artifact seal}
\label{app:repro-artifacts}

This appendix summarizes the sealed evidence package behind the reported numbers. The seal captures the run identifier, source revision, generator seed, metric numerators and denominators, and public artifact digests. It is an integrity record for the synthetic and redacted evidence package, not a release of production source or private runtime traces.

\begin{table}[H]
\centering
\small
\begin{tabular}{ll}
\toprule
Field & Value \\
\midrule
Run id & \texttt{run\_1780535616\_e37d6e1c} \\
Report id & \texttt{scope-aware-memory-1780535616} \\
Source revision & \texttt{7452d12b4111c4367f98fafa67c4af77c8323e40} \\
Generator seed & 42617 \\
Scenario families / scenarios / queries & 13 / 79 / 79 \\
Authorization decision match & 79/79 \\
Permitted-context coverage & 79/79 \\
Forbidden context inclusion & 0/79 \\
Fail-closed coverage & 16/16 \\
Model-neutral assertions & 253/253 \\
Retained supplement digest & \texttt{16146054...} \\
\bottomrule
\end{tabular}
\caption{Durable citation seal for the reported implementation-tier run.}
\label{tab:run-seal}
\end{table}

\begin{table}[H]
\centering
\scriptsize
\begin{tabular}{p{0.52\linewidth}rp{0.18\linewidth}}
\toprule
Public artifact family & Records & SHA-256 prefix \\
\midrule
\texttt{baseline\_report} & 12 & \texttt{f10c9e74} \\
\texttt{model\_neutral\_assertions} & 253 & \texttt{219bbab0} \\
\texttt{authorization\_trace} & 79 & \texttt{74f663fc} \\
\texttt{exact\_context\_snapshot} & 79 & \texttt{3038be23} \\
\texttt{execution\_diagnostics} & 1 & \texttt{a836dbd5} \\
\texttt{expected\_context\_policy} & 79 & \texttt{1554747c} \\
\texttt{synthetic\_input\_events} & 83 & \texttt{3d6dc8ef} \\
\texttt{observation\_mode\_metrics} & 2 & \texttt{8a1a2726} \\
\texttt{provenance\_trace} & 79 & \texttt{4909c4ec} \\
\texttt{unscoped\_baseline\_context\_artifacts} & 79 & \texttt{1ea41aec} \\
\texttt{observed\_context\_artifacts} & 79 & \texttt{749ddc0c} \\
\texttt{scenario\_manifest} & 1 & \texttt{814461ba} \\
\texttt{seeded\_material\_artifacts} & 79 & \texttt{c47bbba6} \\
\texttt{synthetic\_state\_snapshot} & 79 & \texttt{45084623} \\
\texttt{write\_admission\_trace} & 4 & \texttt{f075cba4} \\
\bottomrule
\end{tabular}
\caption{Public artifact families in the retained evidence package. Prefixes are computed over the public projection files listed in \texttt{ARTIFACT\_INDEX.json}.}
\label{tab:artifact-index}
\end{table}

\clearpage

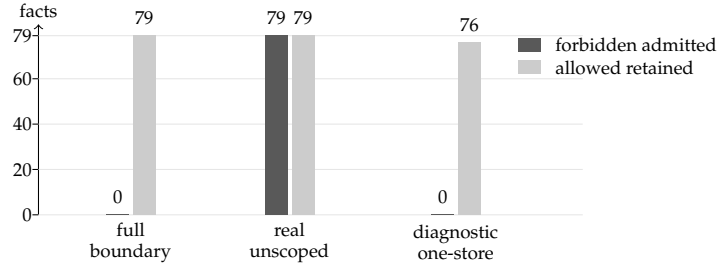
\begin{figure}[H]
\centering
\scriptsize
\begin{tikzpicture}[x=1cm,y=0.03cm]
\draw[->] (0,0) -- (0,84) node[above] {facts};
\foreach \y in {0,20,40,60,79} {
  \draw (-0.08,\y) -- (0,\y) node[left] {\y};
  \draw[black!10] (0,\y) -- (7.2,\y);
}

\fill[black!65] (0.9,0) rectangle (1.2,0);
\node[above] at (1.05,2) {0};
\fill[black!20] (1.25,0) rectangle (1.55,79);
\node[above] at (1.40,81) {79};
\node[align=center] at (1.22,-12) {full\\boundary};

\fill[black!65] (3.0,0) rectangle (3.3,79);
\node[above] at (3.15,81) {79};
\fill[black!20] (3.35,0) rectangle (3.65,79);
\node[above] at (3.50,81) {79};
\node[align=center] at (3.32,-12) {real\\unscoped};

\fill[black!65] (5.2,0) rectangle (5.5,0);
\node[above] at (5.35,2) {0};
\fill[black!20] (5.55,0) rectangle (5.85,76);
\node[above] at (5.70,78) {76};
\node[align=center] at (5.52,-12) {diagnostic\\one-store};

\fill[black!65] (6.35,72) rectangle (6.65,78);
\node[right] at (6.75,75) {forbidden admitted};
\fill[black!20] (6.35,62) rectangle (6.65,68);
\node[right] at (6.75,65) {allowed retained};
\end{tikzpicture}
\caption{Baseline contrast over 79 scenarios. The full boundary retains 79/79 allowed facts while admitting 0/79 forbidden facts. The unscoped real-path baseline retains the same allowed facts but admits 79/79 forbidden facts by construction. The diagnostic one-store projection also admits 0/79 forbidden facts, but retains only 76/79 allowed facts because it omits allowed facts outside that single store; it is not the evaluated full boundary.}
\label{fig:baseline-tradeoff}
\end{figure}

Figure~\ref{fig:baseline-tradeoff} contrasts the evaluated boundary with two baselines in \texttt{baseline\_report}. The unscoped real-path baseline disables authorization and therefore leaks; the one-store group is a synthetic diagnostic projection, not the full method, showing that scoping only one store can preserve confidentiality while dropping allowed context.

\clearpage

\section{Claim-to-evidence closure}
\label{app:claim-evidence}

Each paper claim is tied to machine-readable artifacts rather than to prose-only assertions. Table~\ref{tab:claim-evidence} lists the public artifact families used to close each claim and the metric or structural check that supports it. Claims outside this boundary, including action safety, semantic audience widening, production reliability, and broad memory quality, are excluded or left as future work.

\begin{table}[H]
\centering
\small
\begin{tabular}{p{0.24\linewidth}p{0.34\linewidth}p{0.30\linewidth}}
\toprule
Claim surface & Evidence artifacts & Closure check \\
\midrule
Retrieval and prompt-context boundary & \texttt{authorization\_trace}, \texttt{observed\_context\_artifacts}, \texttt{model\_neutral\_assertions} & 79/79 authorization decisions matched; 0/79 forbidden inclusions. \\
Raw-to-context provenance & \texttt{exact\_context\_snapshot}, \texttt{provenance\_trace} & Every admitted item carries source and audience provenance. \\
Observed source attribution & \texttt{observed\_context\_artifacts}, \texttt{baseline\_report} & Each observed prompt-context item is attributed to a source store; unevaluated memory-quality claims remain excluded. \\
Exact-context supportability & \texttt{exact\_context\_snapshot}, \texttt{provenance\_trace}, \texttt{observed\_context\_artifacts} & The artifacts record what entered the prompt and why it was admitted. \\
Model-neutral invariant & \texttt{model\_neutral\_assertions}, \texttt{authorization\_trace} & 253/253 pre-answer assertions passed without scoring model output. \\
\bottomrule
\end{tabular}
\caption{Claim-to-evidence closure used by the paper.}
\label{tab:claim-evidence}
\end{table}

\clearpage

\section{Worked synthetic scenario}
\label{app:worked-example}

Table~\ref{tab:worked-example} shows selected public artifact fields for one directional cross-channel scenario with the audience boundary disabled and then enabled. Both columns render the same synthetic seed material; the audience boundary is disabled in the unscoped baseline and enabled in the scoped run. For brevity, fact-id references abbreviate the repeated scenario prefix as \texttt{...}.

\begin{table}[H]
\centering
\scriptsize
\begin{tabular}{p{0.23\linewidth}p{0.31\linewidth}p{0.31\linewidth}}
\toprule
Evidence item & Boundary off & Boundary on \\
\midrule
Policy & \begin{tabular}[t]{@{}l@{}}\texttt{required=[...\_group\_fact]} \\ \texttt{forbidden=[...\_alice\_only\_fact]}\end{tabular} & same policy \\
Observed artifact & \texttt{unscoped\_baseline\_context\_artifacts} & \texttt{observed\_context\_artifacts} \\
Authorization & \begin{tabular}[t]{@{}l@{}}\texttt{authorization\_applied=false} \\ \texttt{measurement\_basis=} \\ \texttt{benchmark\_only\_unscoped\_projection}\end{tabular} & \begin{tabular}[t]{@{}l@{}}\texttt{observed\_authorization\_mode=} \\ \texttt{exact\_shared} \\ \texttt{viewer\_contact\_ids=[bob]} \\ \texttt{confidence=high}\end{tabular} \\
Included facts & \begin{tabular}[t]{@{}l@{}}\texttt{included=[...\_group\_fact,} \\ \texttt{\ \ ...\_alice\_only\_fact]}\end{tabular} & \texttt{included=[...\_group\_fact]} \\
Forbidden included & \texttt{[...\_alice\_only\_fact]} & \texttt{[]} \\
\midrule
Allowed seed text & \begin{tabular}[t]{@{}l@{}}\texttt{@@scope\_fact:...\_group\_fact@@} \\ Synthetic directional group fact 1.\end{tabular} & \begin{tabular}[t]{@{}l@{}}\texttt{@@scope\_fact:...\_group\_fact@@} \\ Synthetic directional group fact 1.\end{tabular} \\
Forbidden seed text & \begin{tabular}[t]{@{}l@{}}\texttt{@@scope\_fact:...\_alice\_only\_fact@@} \\ Synthetic directional Alice-only fact 1.\end{tabular} & absent from assembled context \\
\midrule
Exact-context snapshot & not emitted as a separate baseline snapshot row & \begin{tabular}[t]{@{}l@{}}\texttt{fact\_id=...\_group\_fact} \\ no row for \texttt{...\_alice\_only\_fact}\end{tabular} \\
Pre-answer assertions & not applicable to unscoped baseline & \begin{tabular}[t]{@{}l@{}}\texttt{authorization: passed=true} \\ \texttt{forbidden: passed=true} \\ \texttt{allowed: passed=true}\end{tabular} \\
\bottomrule
\end{tabular}
\caption{Boundary-off versus boundary-on evidence for one synthetic scenario. With authorization disabled, the unscoped baseline includes the Alice-only fact in the assembled-context section; with authorization enabled, the observed context includes only the group fact, the forbidden-included set is empty, and the exact-context snapshot contains no row for the forbidden fact.}
\label{tab:worked-example}
\end{table}

\end{document}